\documentclass{amsart}
\usepackage{eurosym}

\theoremstyle{definition}

\theoremstyle{remark}

\numberwithin{equation}{section}
\begin{document}
\title{Classical Symmetries and Painlev\'{e} Analysis for the (1+3)- Kudryashov-Sinelshchikov Equation}
\author{Amlan K Halder}
\address{School of Sciences, Woxsen University, Hyderabad 502345, Telangana, India}
\email{amlanhalder1@gmail.com}
\thanks{}
\author{Rajeswari Seshadri}
\address{Department of Mathematics, Pondicherry University, Puducherry - 605014,
India}
\email{seshadrirajeswari@gmail.com}
\author{A Paliathanasis*}
\address{*Corresponding Author.}
\address{Institute for Systems Science, Durban University of Technology, PO
Box 1334, Durban 4000, South Africa}
\address{School for Data Science and Computational Thinking, Stellenbosch University, Stellenbosch, 7602, South Afrika}
\address{Departamento de Matem\'{a}ticas, Universidad Cat\'{o}lica del Norte, Avda.
Angamos 0610, Casilla 1280 Antofagasta, Chile}
\address{School of Sciences, Woxsen University, Hyderabad 502345, Telangana, India}
\email{anpaliat@phys.uoa.gr}
\thanks{}
\author{PGL Leach}
\address{School of Mathematics, Statistics and Computer Science, University
of KwaZulu-Natal, Durban, South Africa and}
\email{leachp@ukzn.ac.za}
\thanks{}

\date{27:10:2024}

\begin{abstract}
The integrability of the Kudryashov-Sinelshchikov equation in $(1+3)-$ dimension is detected using its point symmetries and the Singularity analysis method. The symmetries mostly points to the existence of infinite-dimensional Lie algebra with the presence of mostly arbitrary functions. The reduction process is initiated by considering these arbitrary functions in the general symmetry vectors. For certain functions, trivial solutions for the general parent equation is obtained whereas for some of the functions, the reductions does not yield an unmitigated result. Finally, a positive result is listed for one of the arbitrary function, for which the singularity analysis method is employed to study the Painlev\'{e} property. The resultant PDE happens to be a fourth-order equation and its approval of the Painlev\'{e} property hints at the integrability of the Kudryashov-Sinelshchikov equation in the general case.
\end{abstract}

\maketitle







\vspace{1.5cc}

\section{Introduction}

In $2012$, Kudryashov and Sinelshchikov \cite{author:Kudryashov} introduced
certain equation discussing the phenomenon of nonlinear waves in the liquid
medium with gas bubbles. The authors cited the importance of the long weakly
nonlinear waves as computationally feasible and efficient. Enormous amount
of work which is focussed on the derivation of exact and special type
solutions have been done by considering the parameters which appears in the
original equation to be of particular value or of functions of the variable $%
t$ \cite{author:Ali,author:El-Shiekh,author:Kumar,author:Seadawy,author:Yang}%
. It is to be noted here that the equation, in general, is considered to be
not integrable, as it fails to pass the necessary requirements of the Painlev
\'{e} test \cite{author:Weiss83a,author:Weiss83b,author:Weiss84}. This
forms the basis of our work. 

In this paper, the symmetries of the equation
by considering the various parameters which appears in the original equation
\cite{author:Kudryashov} are computed and certain reductions are also
listed. Our important observation is one of the reduction with respect to an
arbitrary function symmetry, which leads to a $(1+2)-$ dimensional equation
of fourth-order and is deduced to be integrable by passing the Painlev\'{e}
test. Now, this is quite peculiar, since the order of the parent equation is
also four, this could point to the possibility of integrability for the
parent equation. This result is mentioned in case $6$ and forms the main
result of our work.

The $(1+3)-$ dimensional equation is presented as \cite{author:Kudryashov},
\begin{equation}
\left( \frac{\text{$\mu _{3}$}}{\text{$\mu _{1}$}}+1\right)
u_{xx}u(t,x,y,z)+u_{tx}+\frac{\text{$\mu _{2}$}u_{xxxx}}{2\text{$\mu _{1}$}}%
+\left( \frac{\text{$\mu _{3}$}}{\text{$\mu _{1}$}}+1\right) u_{x}{}^{2}-%
\frac{\text{$\mu _{4}$}u_{xxx}}{2\text{$\mu _{1}$}}+\frac{1}{2}\text{$\mu
_{5}$}^{2}u_{yy}+\frac{1}{2}\text{$\mu _{6}$}u_{zz}=0,  \label{1.1}
\end{equation}%
where $u$ represent the density of the gas bubbles in the liquid medium and
the parameters $\mu _{1},\mu _{2},\mu _{3},\mu _{4},\mu _{5},\mu _{6}$
represent various constant terms describing various physical phenomenon\cite%
{author:El-Shiekh}. The structure of the paper is as follows.

In Section 2 we present the Lie symmetries for the $\left( 1+3\right) $%
-Kudryashov-Sinelshchikov equation. We employ the Lie symmetries to
construct similarity transformations and perform reductions. Then we apply
the singularity analysis to investigate if the reduced equations possess the
Painlev\'{e} property and if the solution can be expressed in terms of
Laurent expansions. In Section 3, we discuss and approach on how to write an
equivalent system which has a finite number of symmetries. We do that by
introducing a new dependent variable, from where we can calculate a
finite-dimensional Lie algebra which leads to the same family of similarity
transformations {}as the infinite-dimensional Lie algebra. Finally, in
Section 4 we draw our conclusions.

\section{Symmetries of Equation (\protect\ref{1.1})}

\label{author:prelim}
The group analysis of differential equations mainly deals with the procedure of obtaining generators of certain transformations which tends to preserve the properties of the differential equations. These generators of transformations represents the underlying algebraic structures of the differential equations under study. Finite representations of these generators leads to proper reductions and possible solutions of the original differential equations and the infinite representations needs proper analysis and treatment to obtain suitable  reductions, for more details we refer the reader to \cite{author:Bluman,author:Olver,author:Meleshko,author:Dimas, author:Dimas2,author:Dimas3,author:Meleshko2}.\\

The general symmetry vector of equation (\ref{1.1}) is
\begin{eqnarray}  \label{2.1}
\Gamma &=& \frac{\text{du} \text{$\mu_1 $} \left(\text{$\mu_5 $}^2 \text{$%
\mu _6 $} \text{b}^{\prime }(t)-\text{$\mu_6$} y \text{c}^{\prime \prime
}(t)-\text{$\mu_5 $}^2 z \text{$d_1$}^{\prime \prime }(t)+\text{$\mu_5 $}^2
y z \text{$d_2$}^{\prime \prime }(t)\right)}{\text{$\mu_5 $}^2 \text{$\mu_6 $%
} (\text{$\mu_1 $}+\text{$\mu_3 $})}  \nonumber \\
&&+\text{dx} \left(\text{b}(t)+\frac{y \left(2 \text{$\mu_5 $}^2 z \text{$d_2
$}^{\prime }(t)-\text{$\mu_6 $} \text{c}^{\prime }(t)\right)}{\text{$\mu_5 $}%
^2 \text{$\mu_6 $}}+\frac{z \left(-\text{$d_1$}^{\prime }(t)-y \text{$d_2$}%
^{\prime }(t)\right)}{\text{$\mu_6 $}}\right)  \nonumber \\
&&+\text{dy} \left(\text{c}(t)-\frac{\text{$\mu_5 $}^2 z \text{$d_2$}(t)}{%
\text{$\mu_6 $}}\right)+\text{dz} (\text{$d_1$}(t)+y \text{$d_2$}(t))+\text{%
dt} \text{$K_{1}$},
\end{eqnarray}
where $K_1$ denote an arbitrary constant and $b(t), c(t), d_{1}(t)$ and $%
d_{2}(t)$ are arbitrary functions with respect to time and hence points to
infinite-dimensional Lie algebra.\newline

\subsection{Reductions of Equation (\protect\ref{1.1})}

\textbf{Case 1}: Using $\partial_{t}$, (with respect to the arbitrary
constant $K_1$) equation (\ref{1.1}) is reduced to
\begin{equation}  \label{2.1a}
\frac{\text{$\mu_3 $} v_{xx} v(x,y,z)}{\text{$\mu_1 $}}+v_{xx} v(x,y,z)+%
\frac{\text{$\mu_2 $} v_{xxxx}}{2 \text{$\mu_1 $}}+\frac{\text{$\mu_3 $}
v_{x}{}^2}{\text{$\mu_1 $}}-\frac{\text{$\mu_4 $} v_{xxx}}{2 \text{$\mu_1 $}}%
+v_{x}{}^2+\frac{1}{2} \text{$\mu_5 $}^2 v_{yy}+\frac{1}{2} \text{$\mu_6 $}
v_{zz}=0.
\end{equation}
The equation (\ref{2.1a}) is a PDE in $(1+2)$ dimension.\newline
The order of the reduced equation is similar to that of the parent equation.
The analysis of this equation is under study and there is a possibility that
the obtained symmetries could be similar to that of the parent equation and
also certain chances of obtaining some new symmetries which are the hidden
symmetries of the original equation.\newline

\textbf{Case 2}: Using the symmetry vector with respect to the arbitrary
function $d_{1}(t)$, we obtain
\begin{equation}  \label{2.2a}
\frac{1}{2} \text{$\mu_5 $}^2 v_{yy}-\frac{\text{$\mu_1 $} d_{1}^{\prime
\prime }(t)}{2 d_{1}(t) (\text{$\mu_1 $}+\text{$\mu_3 $})}=0.
\end{equation}
The similarity varaiable is
\begin{equation}  \label{2.3a}
u(t,x,y,z)=v(t,y)-\frac{\text{$\mu_1 $} z^2 d_{1}^{\prime \prime }(t)}{2
\text{$\mu _6$} d_{1}(t) (\text{$\mu _1$}+\text{$\mu_3 $})}.
\end{equation}

Equation (\ref{2.2a}) is easily solvable and hence points to the existence
of trivial solution for equation (\ref{1.1}).\newline

\textbf{Case 3}: Using another similarity variable with respect to $d_{1}(t)$%
, we obtain
\begin{equation}  \label{2.4a}
\text{$\mu_5 $}^2 d_{1}^{\prime}(t) (\text{$\mu_1 $}+\text{$\mu_3 $})
v_{yy}+2 \text{$\mu_1 $} d_{1}^{\prime \prime \prime }(t)=0.
\end{equation}
The similarity variable is
\begin{equation}  \label{2.5a}
u(t,x,y,z)=v(t,y)+\frac{\text{$\mu_1 $} x d_{1}^{\prime \prime }(t)}{%
d_{1}^{\prime}(t) (\text{$\mu_1 $}+\text{$\mu_3 $})}.
\end{equation}

Equation (\ref{2.4a}) is also solvable and hence provides trivial solution
to equation (\ref{1.1}).\newline

\textbf{Case 4}: Here, we list certain similarity transformations with
respect to the symmetry vector $d_{2}(t)$, which does not yield any positive
result.
\begin{eqnarray}  \label{2.6a}
u(t,x,y,z)&=&v\left(t,\text{$\mu_6 $} y^2+\text{$\mu_5 $}^2 z^2,\frac{y
d_{2}^{\prime }}{d_{2}(t)}+\text{$\mu_5 $}^2 x\right)-\frac{\text{$\mu_1 $}
y^2 d_{2}^{\prime \prime }(t)}{2 \text{$\mu_5 $}^2 d_{2}(t) (\text{$\mu_1 $}+%
\text{$\mu_3 $})},  \nonumber \\
u(t,x,y,z)&=&v\left(t,\text{$\mu_6 $} y^2+\text{$\mu_5 $}^2 z^2,\frac{z^2
d_{2}^{\prime }}{2 d_{2}(t)}-\text{$\mu_6 $} x\right)-\frac{\text{$\mu_1 $}
y^2 d_{2}^{\prime \prime }(t)}{2 \text{$\mu_5 $}^2 d_{2}(t) (\text{$\mu_1 $}+%
\text{$\mu_3 $})},  \nonumber \\
u(t,x,y,z)&=&v\left(t,\frac{y d_{2}^{\prime }}{d_{2}(t)}+\text{$\mu_5 $}^2 x,%
\frac{z^2 d_{2}^{\prime }}{2 d_{2}(t)}-\text{$\mu_6 $} x\right)-\frac{\text{$%
\mu_1 $} y^2 d_{2}^{\prime \prime }(t)}{2 \text{$\mu_5 $}^2 d_{2}(t) (\text{$%
\mu_1 $}+\text{$\mu_3 $})},  \nonumber \\
u(t,x,y,z)&=&v\left(t,\text{$\mu_6 $} y^2+\text{$\mu_5 $}^2 z^2,\frac{y {%
d_{2}}^{\prime }(t)}{d_{2}(t)}+\text{$\mu_5$}^2 x\right)+\frac{\text{$\mu_1 $%
} z^2 d_{2}^{\prime \prime }(t)}{2 \text{$\mu_6 $} d_{2}(t) (\text{$\mu_1 $}+%
\text{$\mu_3 $})},  \nonumber \\
u(t,x,y,z)&=&v\left(t,\text{$\mu_6 $} y^2+\text{$\mu_5 $}^2 z^2,\frac{y {%
d_{2}}^{\prime }(t)}{d_{2}(t)}+\text{$\mu_5 $}^2 x\right)+\frac{\text{$\mu_1
$} x d_{2}^{\prime \prime }(t)}{{d_{2}}^{\prime }(t) (\text{$\mu_1 $}+\text{$%
\mu_3 $})},  \nonumber \\
u(t,x,y,z)&=&v\left(t,\text{$\mu_6 $} y^2+\text{$\mu_5 $}^2 z^2,\frac{z^2 {%
d_{2}}^{\prime }(t)}{2 d_{2}(t)}-\text{$\mu_6 $} x\right)+\frac{\text{$\mu_1
$} x d_{2}^{\prime \prime }(t)}{{d_{2}}^{\prime }(t) (\text{$\mu_1 $}+\text{$%
\mu_3 $})},  \nonumber \\
u(t,x,y,z)&=&v\left(t,\frac{y {d_{2}}^{\prime }(t)}{d_{2}(t)}+\text{$\mu_5 $}%
^2 x,\frac{z^2 {d_{2}}^{\prime }(t)}{2 d_{2}(t)}-\text{$\mu_6 $} x\right)+%
\frac{\text{$\mu_1 $} x d_{2}^{\prime \prime }(t)}{{d_{2}}^{\prime }(t) (%
\text{$\mu_1 $}+\text{$\mu_3 $})}.
\end{eqnarray}

\textbf{Case 5}: Next, we mention the symmetry reduction with respect to
traveling-wave reduction for equation (\ref{1.1}). For a substitution of $%
u(t,x,y,z)= v(k x+l y+m z-c t), s=k x+l y+m z-c t$, the equation obtained is
as follows:

\begin{eqnarray}  \label{2.7a}
-c k v^{\prime \prime }(s)+\frac{k^4 \text{$\mu_2 $} v^{\prime \prime \prime
\prime }(s)}{2 \text{$\mu_1 $}}-\frac{k^3 \text{$\mu_4 $} v^{\prime \prime
\prime }(s)}{2 \text{$\mu_1 $}}+\frac{k^2 \text{$\mu_3 $} v(s) v^{\prime
\prime }(s)}{\text{$\mu_1 $}}+k^2 v(s) v^{\prime \prime }(s)+\frac{k^2 \text{%
$\mu_3 $} {v^{\prime }(s)}^2}{\text{$\mu_1 $}}+k^2 {v^{\prime }(s)}^2
\nonumber \\
+\frac{1}{2} l^2 \text{$\mu_5 $}^2 v^{\prime \prime }(t)+\frac{1}{2} \text{$%
\mu_6 $} m^2 v^{\prime \prime }(s)&=&0.  \nonumber \\
\end{eqnarray}

As equation (\ref{2.7a}) is an autonomous equation, the sole symmetry vector
is $\partial_{s}$, which reduces it to a third-order equation which are
devoid of any point symmetries.

\begin{eqnarray}  \label{2.8a}
p^{\prime \prime \prime }(q)&=&-\frac{p(q)^2 p^{\prime }(q) \left(-2 c k
\text{$\mu_1 $}+2 k^2 \text{$\mu_1 $} q+2 k^2 \text{$\mu_3 $} q+l^2 \text{$%
\mu_1 $} \text{$\mu_5 $}^2+\text{$\mu_1 $} \text{$\mu_6 $} \text{m}^2\right)%
}{k^4 \text{$\mu_2 $}}+\frac{(2 \text{$\mu_1 $}+2 \text{$\mu_3 $}) p(q)^3}{%
k^2 \text{$\mu_2 $}}-\frac{3 \text{$\mu_4 $} {p^{\prime }(q)}^2}{k \text{$%
\mu_2 $}}  \nonumber \\
&&+p^{\prime \prime }(q) \left(\frac{\text{$\mu_4 $} p(q)}{k \text{$\mu_2 $}}%
+\frac{10 p^{\prime }(q)}{p(q)}\right)-\frac{15 {p^{\prime }(q)}^3}{p(q)^2},
\end{eqnarray}
where $q=v(s)$ and $p(q)=\frac{1}{v^{\prime }(s)}$. To determine the
integrability of equation (\ref{2.8a}), the method of singularity analysis
is used.  The method basically consists of verification of suitability for the differential equation to pass the Painlev\'{e} property. The objective of this method consists of looking for singularities, which are in the form of poles and then deriving series solutions, with positive or negative terms or in some cases a combination of both for the differential equation under study. \cite{author:Leach,author:Conte1,author:Conte2,author:ARS1, author:ARS2, author:ARS3, author:Paliathanasis}\\  

The equation (\ref{2.8a}) is rewritten suitably in the new form $%
y(x)$, which leads to

\begin{eqnarray}  \label{2.9a}
-2 c k \text{$\mu_1 $} y(x)^4 y^{\prime 4 }\text{$\mu_2 $} y(x)^2 y^{\prime
\prime \prime }-10 k^4 \text{$\mu_2 $} y(x) y^{\prime \prime \prime }+15 k^4
\text{$\mu_2 $2} y^{\prime 3}-k^3 \text{$\mu_4 $} y(x)^3 y^{\prime \prime
}+3 k^3 \text{$\mu_4 $} y(x)^2 y^{\prime 2}  \nonumber \\
+2 k^2 \text{$\mu_1 $} x y(x)^4 y^{\prime 2 }\text{$\mu_3 $} x y(x)^4
y^{\prime 2 }\text{$\mu_1 $} y(x)^5-2 k^2 \text{$\mu_3 $} y(x)^5+l^2 \text{$%
\mu_1 $} \text{$\mu_5 $}^2 y(x)^4 y^{\prime }+\text{$\mu_1 $} \text{$\mu_6 $}
\text{m}^2 y(x)^4 y^{\prime }&=&0.  \nonumber \\
\end{eqnarray}

To detect the existence of movable singularity, $y\rightarrow \alpha
(x-x_{0})$, is substituted in equation (\ref{2.9a}), which leads to
\begin{eqnarray}  \label{2.10a}
-2 \alpha ^5 c k \text{$\mu_1 $} p z^{5 p-1}+6 \alpha ^3 k^4 \text{$\mu_2 $}
p^3 z^{3 p-3}+7 \alpha ^3 k^4 \text{$\mu_2 $} p^2 z^{3 p-3}+2 \alpha ^3 k^4
\text{$\mu_2 $} p z^{3 p-3}+2 \alpha ^4 k^3 \text{$\mu_4 $} p^2 z^{4 p-2}
\nonumber \\
+\alpha ^4 k^3 \text{$\mu_4$} p z^{4 p-2} +2 \alpha^{5} k^{2} \text{$\mu_1 $}
px_{0} z^{5p-1}+2 \alpha^{5} k^{2} \text{$\mu_3$} p x_{0} z^{5 p-1} -2
\alpha^5 k^2 \text{$\mu_1 $} z^{5 p}+2 \alpha ^5 k^2 \text{$\mu_1 $} p z^{5
p}  \nonumber \\
-2 \alpha ^5 k^2 \text{$\mu_3$} z^{5 p}+2 \alpha ^5 k^2 \text{$\mu_3 $} p
z^{5 p}+\alpha ^5 l^2 \text{$\mu_1 $} \text{$\mu_5$}^2 p z^{5 p-1} +\alpha ^5%
\text{$\mu_1 $} \text{$\mu_6 $} \text{m}^2 p z^{5 p-1}.  \nonumber \\
\end{eqnarray}
By considering the dominating terms, following values of the exponent $p$
are obtained $(-2,-\frac{3}{2},-1)$. The initial computation is started by
considering the value of $p$ to be $-1$ which leads to:
\begin{eqnarray}  \label{2.11a}
\frac{2 \alpha ^5 c k \text{$\mu_1 $}}{z^6}-\frac{\alpha ^3 k^4 \text{$\mu_2
$}}{z^6}+\frac{\alpha ^4 k^3 \text{$\mu_4 $}}{z^6}-\frac{2 \alpha ^5 k^2
\text{$\mu_1 $} \text{x0}}{z^6}-\frac{2 \alpha ^5 k^2 \text{$\mu_3 $} x_{0}}{%
z^6}-\frac{4 \alpha ^5 k^2 \text{$\mu_1 $}}{z^5}-\frac{4 \alpha ^5 k^2 \text{%
$\mu_3 $}}{z^5}  \nonumber \\
-\frac{\alpha ^5 l^2 \text{$\mu_1 $} \text{$\mu_5 $}^2}{z^6}-\frac{\alpha ^5
\text{$\mu_1 $} \text{$\mu_6 $} \text{m}^2}{z^6}
\end{eqnarray}
By taking into account the dominating terms, following values of the
leading-order coefficient are obtained.
\[
\alpha \rightarrow 0(3),
\]

\[
\alpha  \rightarrow \frac{-k^3\text{$\mu_4 $}-k^2 \sqrt{8 c k \text{$\mu_1 $}
\text{$\mu_2 $}+k^2 \text{$\mu_4 $}^2-8 k^2 \text{$\mu_1 $} \text{$\mu_2 $}
x_{0}-8 k^2 \text{$\mu_2 $} \text{$\mu_3 $} x_{0}-4 l^2 \text{$\mu_1 $}
\text{$\mu_2 $} \text{$\mu_5 $}^2-4 \text{$\mu_1 $} \text{$\mu_2 $} \text{$%
\mu_6 $} \text{m}^2}}{2 \left(2 c k \text{$\mu_1 $}-2 k^2 \text{$\mu_1 $}
x_{0}-2 k^2 \text{$\mu_3 $} x_{0}-l^2 \text{$\mu_1 $} \text{$\mu_5 $}^2-%
\text{$\mu_1 $} \text{$\mu_6 $}\text{m}^2\right)},
\]
\[
\alpha \rightarrow \frac{-k^3 \text{$\mu_4 $}+k^2 \sqrt{8 c k \text{$\mu_1 $}
\text{$\mu_2 $}+k^2 \text{$\mu_4 $}^2-8 k^2 \text{$\mu_1 $} \text{$\mu_2 $}
x_{0}-8 k^2 \text{$\mu_2 $} \text{$\mu_3 $} x_{0}-4 l^2 \text{$\mu_1 $}
\text{$\mu_2 $} \text{$\mu_5 $}^2-4 \text{$\mu_1 $} \text{$\mu_2 $} \text{$%
\mu_6 $} \text{m}^2}}{2 \left(2 c k \text{$\mu_1 $}-2 k^2 \text{$\mu_1 $}
x_{0}-2 k^2 \text{$\mu_3 $} x_{0}-l^2 \text{$\mu_1 $} \text{$\mu_5 $}^2-%
\text{$\mu_1 $} \text{$\mu_6 $} \text{m}^2\right)}.
\]

Next to compute the resonances the following substitution is employed in
equation (\ref{2.9a})
\[
y\rightarrow \alpha (x-x_{0})^{(-1)}+m_{1}(x-x_{0})^{(-1+s)}.
\]
The coefficient of $m_{1}$ is collected and for a nonzero value of $\alpha$
the following values of resonance $s$ are obtained.\newline
\[
s\rightarrow -1(2),
\]
\begin{eqnarray*}
s&=& \frac{1}{4 c k \text{$\mu_1 $} \text{$\mu_2 $}-4 k^2 \text{$\mu_1 $}
\text{$\mu_2 $} x_{0}-4 k^2 \text{$\mu_2 $} \text{$\mu_3 $} x_{0}-2 l^2
\text{$\mu_1 $} \text{$\mu_2 $} \text{$\mu_5 $}^2-2 \text{$\mu_1 $} \text{$%
\mu_2 $} \text{$\mu_6 $} \text{m}^2}\left(A_{\nu}\right),
\end{eqnarray*}
where,
\begin{eqnarray*}
A_{\nu}&=&\left(
\begin{array}{c}
-k \text{$\mu_4 $} \sqrt{8 c k \text{$\mu_1 $} \text{$\mu_2 $}+k^{2} \left(%
\text{$\mu_4 $}^{2}-8 \text{$\mu_2 $} x_{0} (\text{$\mu_1 $}+\text{$\mu_3 $}%
)\right)-4 \text{$\mu_1 $} \text{$\mu_2 $} \left(l^{2} \text{$\mu_5 $}^{2}+%
\text{$\mu_6 $} \text{m}^{2}\right)}-8 c k \text{$\mu_1 $} \text{$\mu_2 $}%
-k^{2} \text{$\mu_4 $}^{2} \\
+8 k^2 \text{$\mu_1 $} \text{$\mu_2 $} x_{0}+8 k^2 \text{$\mu_2 $} \text{$%
\mu_3 $} x_{0}+4 l^2 \text{$\mu_1 $} \text{$\mu_2 $} \text{$\mu_5 $}^2+4
\text{$\mu_1 $} \text{$\mu_2 $} \text{$\mu_6 $} \text{m}^2%
\end{array}
\right).
\end{eqnarray*}

\textbf{\textit{Observations}}: The occurrence of $-1$ values for the
resonances are associated with the arbitrary location of the singularity $%
x_{0}$. Interestingly, the other value for the resonance is dependent on the
equation parameters, wave numbers, frequency and most importantly to the
location of the singularity $x_{0}$. Such cases (dependence on $x_{0}$) are
observed very few in the literature. Currently, those specific cases are
under study.\newline

\textbf{Case 6}: Reduction using the symmetry vector $c(t)$.

Using the similarity transformation
\begin{eqnarray}  \label{2.12a}
u(t,x,y,z)&=&v\left(t,z,\frac{y^2 c^{\prime }(t)}{2 c(t)}+\text{$\mu_5 $}^2
x\right)-\frac{\text{$\mu_1 $} y^2 c^{\prime \prime }(t)}{2 \text{$\mu_5 $}%
^2 c(t) (\text{$\mu_1 $}+\text{$\mu_3 $})},
\end{eqnarray}
the equation (\ref{1.1}) reduces to
\begin{eqnarray}  \label{2.13a}
0&=&-\text{$\mu_1 $}^2 c^{\prime \prime }(t)+ (\text{$\mu_1 $}+\text{$\mu_3 $%
}) \left( \text{$\mu_1 $} \text{$\mu_5 $}^2 c^{\prime }(t) v_{\nu }+c(t) %
\left[
\begin{array}{c}
2 \text{$\mu_5 $}^4 (\text{$\mu_1 $}+\text{$\mu_3 $}) v_{\nu \nu } v(t,z,\nu
)+2 \text{$\mu_1 $} \text{$\mu_5 $}^2 v_{t\nu } \\
+2 \text{$\mu_5 $}^4 (\text{$\mu_1 $}+\text{$\mu_3 $}) v_{\nu }{}^2+\text{$%
\mu_2 $} \text{$\mu_5 $}^8 v_{\nu \nu \nu \nu }-\text{$\mu_4 $} \text{$\mu_5
$}^6 v_{\nu \nu \nu } \\
+\text{$\mu_1 $} \text{$\mu_6 $} v_{zz}%
\end{array}
\right] \right).  \nonumber \\
\end{eqnarray}
Computation of determining equations is difficult for the fourth-order
equation in $(1+2)-$ dimension for the generic term $c(t)$. Though, we must
specify here that for certain particular cases of $c(t)$, such as the power
functions, certain notes pointing to nicer results are obtained. This arises
a question of the integrability of the equation in general. Hence, the
Singularity analysis for PDEs being employed to study its integrability.%
\newline

To begin with, a substitution of $v\rightarrow V_{0}\phi(t,z,\nu)^{p}$ is
done, where $p$ denotes the leading-order exponent. The substitution leads
to
\begin{eqnarray}  \label{2.14a}
-p V_{0} (\text{$\mu_1 $}+\text{$\mu_3 $}) \phi (t,z,\nu )^{p-4} \left(
\begin{array}{c}
-\phi (t,z,\nu )^3 \left(\text{$\mu_1 $} \text{$\mu_5 $}^2 c^{\prime }\phi
_{\nu }+c(t) A_{\nu_{1}}\right) +(p-1) c(t) \phi (t,z,\nu )^2 \left(
A_{\nu_{2}} \right) \nonumber \\
+\text{$\mu_5 $}^6 (p-2) (p-1) c(t) \phi _{\nu }{}^2 \phi (t,z,\nu ) \left(%
\text{$\mu_4 $} \phi _{\nu }-6 \text{$\mu_2 $} \text{$\mu_5 $}^2 \phi _{\nu
\nu }\right)\nonumber \\
+\text{$\mu_2 $} \text{$\mu_5 $}^8 (-(p-3)) (p-2) (p-1) c(t) \phi _{\nu }{}^4%
\end{array}
\right)  \nonumber \\
+2 \text{$\mu_5 $}^4 p V_{0}^2 c(t) (\text{$\mu_1 $}+\text{$\mu_3 $})^2 \phi
(t,z,\nu )^{2 p-2} \left(\phi _{\nu \nu } \phi (t,z,\nu )+(2 p-1) \phi _{\nu
}{}^2\right)-\text{$\mu_1 $}^2 c^{\prime \prime }(t),  \nonumber \\
\end{eqnarray}
where,
\begin{eqnarray*}
A_{\nu_{1}}&=&\left(
\begin{array}{c}
\text{$\mu_2 $} \text{$\mu_5 $}^8 \phi _{\nu \nu \nu \nu }-\text{$\mu_4 $}
\text{$\mu_5 $}^6 \phi _{\nu \nu \nu }+2 \text{$\mu_1 $} \text{$\mu_5 $}^2
\phi _{t\nu } \\
+\text{$\mu_1 $} \text{$\mu_6 $} \phi _{zz}%
\end{array}
\right),
\end{eqnarray*}
and
\begin{eqnarray*}
A_{\nu_{2}}&=&\left(
\begin{array}{c}
-3 \text{$\mu_2 $} \text{$\mu_5 $}^8 \phi _{\nu \nu }{}^2+\phi _{\nu }
\left(-4 \text{$\mu_2 $} \text{$\mu_5 $}^8 \phi _{\nu \nu \nu }+3 \text{$%
\mu_4 $} \text{$\mu_5 $}^6 \phi _{\nu \nu }-2 \text{$\mu_1 $} \text{$\mu_5 $}%
^2 \phi _{t}\right)-\text{$\mu_1 $} \text{$\mu_6 $} \phi _{z}{}^2%
\end{array}
\right).
\end{eqnarray*}
One of the possible values of the leading-order exponent is $-2$. The
subsequent dominating terms are
\begin{eqnarray}  \label{2.15a}
20 \text{$\mu_1 $}^2 \text{$\mu_5 $}^4 V_{0}^2 c(t) \phi _{\nu }{}^2+40
\text{$\mu_1 $} \text{$\mu_3 $} \text{$\mu_5 $}^4 V_{0}^2 c(t) \phi _{\nu
}{}^2+20 \text{$\mu_3 $}^2 \text{$\mu_5 $}^4 V_{0}^2 c(t) \phi _{\nu
}{}^2+120 \text{$\mu_1 $} \text{$\mu_2 $} \text{$\mu_5 $}^8 V_{0} c(t) \phi
_{\nu }{}^4  \nonumber \\
+120 \text{$\mu_2 $} \text{$\mu_3 $} \text{$\mu_5 $}^8 V_{0} c(t) \phi _{\nu
}{}^4  \nonumber \\
\end{eqnarray}

After factorizing, the possible value of the leading-order coefficient $V_{0}
$ is
\[
\left\{V_{0}\to  0, V_{0}\to -\frac{6 \text{$\mu_2 $} \text{$\mu_5 $}^4 \phi
_{\nu }{}^2}{\text{$\mu_1 $}+\text{$\mu_3 $}}\right\}
\]

Next, to compute the resonances, the following substitution is made:
\[
m \phi (t,z,\nu  )^{S-2}+\frac{V_{0}}{\phi (t,z,\nu )^2}
\]

From the substituted equation, the coefficients of linear terms in $m$ are
collected and hence the values of the resonances are computed. The values
are
\[
S\rightarrow -1, 4,5,6.
\]
To verify the consistency of the obtained values, the following truncated
Laurent series solution is substituted in the reduced equation with respect
to $c(t)$. The objective is to obtain the required number of arbitrary
constants.
\begin{eqnarray}
v(t,z,\nu)&=&-\frac{6 \text{$\mu_2 $} \text{$\mu_5 $}^4 \phi _{\nu }{}^2}{(%
\text{$\mu_1 $}+\text{$\mu_3 $}) \phi (t,z,\nu )^2}+\frac{V_{1}(t,z,\nu )}{%
\phi (t,z,\nu )}+V_{2}(t,z,\nu )+V_{3}(t,z,\nu ) \phi (t,z,\nu
)+V_{4}(t,z,\nu ) \phi (t,z,\nu )^2  \nonumber \\
&&+V_{5}(t,z,\nu ) \phi (t,z,\nu )^3+V_{6}(t,z,\nu ) \phi (t,z,\nu
)^4+V_{7}(t,z,\nu ) \phi (t,z,\nu )^5+V_{8}(t,z,\nu ) \phi (t,z,\nu )^6
\nonumber \\
\end{eqnarray}

By collecting the terms, it is observed that $V_{1},V_{2},V_{3},V_{7},V_{8}$
are fixed and $V_{4},V_{5},V_{6}$ are arbitrary.\newline
The coefficient of $V_{1}(t,z,\nu )$ is
\[
V_{1}(t,z,\nu )=\frac{6\text{$\mu _{5}$}^{2}\left( 5\text{$\mu _{2}$}\text{$%
\mu _{5}$}^{2}\phi _{\nu \nu }-\text{$\mu _{4}$}\phi _{\nu }\right) }{5(%
\text{$\mu _{1}$}+\text{$\mu _{3}$})}.
\]%
The coefficient of $V_{2}(t,z,\nu )$ is
\[
V_{2}(t,z,\nu )=\frac{1}{12\text{$\mu _{2}$}\text{$\mu _{5}$}^{4}(\text{$\mu
_{1}$}+\text{$\mu _{3}$})\phi _{\nu }{}^{2}}\left(
\begin{array}{c}
24\text{$\mu _{1}$}\text{$\mu _{2}$}\text{$\mu _{5}$}^{4}\phi _{\nu \nu
}V_{1}+\text{$\mu _{1}$}\text{$\mu _{4}$}\text{$\mu _{5}$}^{2}\phi _{\nu
}V_{1} \\
+24\text{$\mu _{2}$}\text{$\mu _{3}$}\text{$\mu _{5}$}^{4}\phi _{\nu \nu
}V_{1}+\text{$\mu _{3}$}\text{$\mu _{4}$}\text{$\mu _{5}$}^{2}\phi _{\nu
}V_{1} \\
+\text{$\mu _{1}$}^{2}{V_{1}}^{2}+2\text{$\mu _{1}$}\text{$\mu _{3}$}{V_{1}}%
^{2}+\text{$\mu _{3}$}^{2}{V_{1}}^{2} \\
-162\text{$\mu _{2}$}^{2}\text{$\mu _{5}$}^{8}\phi _{\nu \nu }{}^{2}-72\text{%
$\mu _{2}$}^{2}\text{$\mu _{5}$}^{8}\phi _{\nu }\phi _{\nu \nu \nu }+54\text{%
$\mu _{2}$}\text{$\mu _{4}$}\text{$\mu _{5}$}^{6}\phi _{\nu }\phi _{\nu \nu }
\\
-12\text{$\mu _{1}$}\text{$\mu _{2}$}\text{$\mu _{5}$}^{2}\phi _{\nu }\phi
_{t}+8\text{$\mu _{1}$}\text{$\mu _{2}$}\text{$\mu _{5}$}^{4}\phi _{\nu }{%
V_{1}}_{\nu }+8\text{$\mu _{2}$}\text{$\mu _{3}$}\text{$\mu _{5}$}^{4}\phi
_{\nu }{V_{1}}_{\nu } \\
-6\text{$\mu _{1}$}\text{$\mu _{2}$}\text{$\mu _{6}$}\phi _{z}{}^{2}%
\end{array}%
\right) .
\]%
The coefficient of $V_{3}(t,z,\nu )$ is
\[
V_{3}(t,z,\nu )=\frac{1}{12\text{$\mu _{2}$}(\text{$\mu _{1}$}+\text{$\mu
_{3}$})\text{$\mu _{5}$}^{8}c(t)\phi _{\nu }{}^{4}}\left(
\begin{array}{c}
72\text{$\mu _{2}$}^{2}c(t)\phi _{\nu \nu }{}^{3}\text{$\mu _{5}$}^{12}+264%
\text{$\mu _{2}$}^{2}c(t)\phi _{\nu }\phi _{\nu \nu }\phi _{\nu \nu \nu }%
\text{$\mu _{5}$}^{12} \\
+54\text{$\mu _{2}$}^{2}c(t)\phi _{\nu }{}^{2}\phi _{\nu \nu \nu \nu }\text{$%
\mu _{5}$}^{12}-72\text{$\mu _{2}$}\text{$\mu _{4}$}c(t)\phi _{\nu }\phi
_{\nu \nu }{}^{2}\text{$\mu _{5}$}^{10} \\
-42\text{$\mu _{2}$}\text{$\mu _{4}$}c(t)\phi _{\nu }{}^{2}\phi _{\nu \nu
\nu }\text{$\mu _{5}$}^{10}+24\text{$\mu _{1}$}\text{$\mu _{2}$}c(t){V_{2}}%
_{\nu }\phi _{\nu }{}^{3}\text{$\mu _{5}$}^{8} \\
+24\text{$\mu _{2}$}\text{$\mu _{3}$}c(t){V_{2}}_{\nu }\phi _{\nu }{}^{3}%
\text{$\mu _{5}$}^{8}-9\text{$\mu _{1}$}\text{$\mu _{2}$}c(t){V_{1}}\phi
_{\nu \nu }{}^{2}\text{$\mu _{5}$}^{8} \\
-9\text{$\mu _{2}$}\text{$\mu _{3}$}c(t){V_{1}}\phi _{\nu \nu }{}^{2}\text{$%
\mu _{5}$}^{8}+60\text{$\mu _{1}$}\text{$\mu _{2}$}c(t){V_{2}}\phi _{\nu
}{}^{2}\phi _{\nu \nu }\text{$\mu _{5}$}^{8} \\
+60\text{$\mu _{2}$}\text{$\mu _{3}$}c(t){V_{2}}\phi _{\nu }{}^{2}\phi _{\nu
\nu }\text{$\mu _{5}$}^{8}-12\text{$\mu _{1}$}\text{$\mu _{2}$}c(t){V_{1}}%
_{\nu }\phi _{\nu }\phi _{\nu \nu }\text{$\mu _{5}$}^{8} \\
-12\text{$\mu _{2}$}\text{$\mu _{3}$}c(t){V_{1}}_{\nu }\phi _{\nu }\phi
_{\nu \nu }\text{$\mu _{5}$}^{8}-8\text{$\mu _{1}$}\text{$\mu _{2}$}c(t){%
V_{1}}\phi _{\nu }\phi _{\nu \nu \nu }\text{$\mu _{5}$}^{8} \\
-8\text{$\mu _{2}$}\text{$\mu _{3}$}c(t){V_{1}}\phi _{\nu }\phi _{\nu \nu
\nu }\text{$\mu _{5}$}^{8}+6\text{$\mu _{1}$}\text{$\mu _{2}$}c^{\prime
}(t)\phi _{\nu }{}^{3}\text{$\mu _{5}$}^{6} \\
-3\text{$\mu _{1}$}\text{$\mu _{4}$}c(t){V_{1}}_{\nu }\phi _{\nu }{}^{2}%
\text{$\mu _{5}$}^{6}-3\text{$\mu _{3}$}\text{$\mu _{4}$}c(t){V_{1}}_{\nu
}\phi _{\nu }{}^{2}\text{$\mu _{5}$}^{6} \\
-3\text{$\mu _{1}$}\text{$\mu _{4}$}c(t){V_{1}}\phi _{\nu }\phi _{\nu \nu }%
\text{$\mu _{5}$}^{6}-3\text{$\mu _{3}$}\text{$\mu _{4}$}c(t){V_{1}}\phi
_{\nu }\phi _{\nu \nu }\text{$\mu _{5}$}^{6} \\
+24\text{$\mu _{1}$}\text{$\mu _{2}$}c(t)\phi _{\nu }\phi _{\nu \nu }\phi
_{t}\text{$\mu _{5}$}^{6}+36\text{$\mu _{1}$}\text{$\mu _{2}$}c(t)\phi _{\nu
}{}^{2}\phi _{t\nu }\text{$\mu _{5}$}^{6} \\
+2\text{$\mu $1}^{2}c(t)V_{1}V_{2}\phi _{\nu }{}^{2}\text{$\mu _{5}$}^{4}+2%
\text{$\mu _{3}$}^{2}c(t)V_{1}V_{2}\phi _{\nu }{}^{2}\text{$\mu _{5}$}^{4}
\\
+4\text{$\mu _{1}$}\text{$\mu _{3}$}c(t)V_{1}V_{2}\phi _{\nu }{}^{2}\text{$%
\mu _{5}$}^{4}-4\text{$\mu _{1}$}^{2}c(t)V_{1}{V_{1}}_{\nu }\phi _{\nu }%
\text{$\mu _{5}$}^{4} \\
-4\text{$\mu _{3}$}^{2}c(t)V_{1}{V_{1}}_{\nu }\phi _{\nu }\text{$\mu _{5}$}%
^{4}-8\text{$\mu _{1}$}\text{$\mu _{3}$}c(t)V_{1}{V_{1}}_{\nu }\phi _{\nu }%
\text{$\mu _{5}$}^{4} \\
-\text{$\mu _{1}$}^{2}c(t)V_{1}^{2}\phi _{\nu \nu }\text{$\mu _{5}$}^{4}-%
\text{$\mu _{3}$}^{2}c(t)V_{1}^{2}\phi _{\nu \nu }\text{$\mu _{5}$}^{4}-2%
\text{$\mu _{1}$}\text{$\mu _{3}$}c(t)V_{1}^{2}\phi _{\nu \nu }\text{$\mu
_{5}$}^{4} \\
+24\text{$\mu _{1}$}\text{$\mu _{2}$}\text{$\mu _{6}$}c(t)\phi _{\nu }\phi
_{z}\phi _{z\nu }\text{$\mu _{5}$}^{4}+6\text{$\mu _{1}$}\text{$\mu _{2}$}%
\text{$\mu _{6}$}c(t)\phi _{\nu }{}^{2}\phi _{zz}\text{$\mu _{5}$}^{4} \\
+2\text{$\mu _{1}$}^{2}c(t)V_{1}\phi _{\nu }\phi _{t}\text{$\mu _{5}$}^{2}+2%
\text{$\mu _{1}$}\text{$\mu _{3}$}c(t)V_{1}\phi _{\nu }\phi _{t}\text{$\mu
_{5}$}^{2} \\
+\text{$\mu _{1}$}^{2}\text{$\mu _{6}$}c(t)V_{1}\phi _{z}{}^{2}+\text{$\mu
_{1}$}\text{$\mu _{3}$}\text{$\mu _{6}$}c(t)V_{1}\phi _{z}{}^{2}%
\end{array}%
\right) .
\]%
Hence, $V_{4}(t,z,\nu ),V_{5}(t,z,\nu ),V_{6}(t,z,\nu )$ are arbitrary and
therefore the truncated series solutions points to the existence of complete
integrability for equation (\ref{2.13a}).

\section{From infinity to a finite Lie algebra}

We found before that equation (\ref{1.1}) admits an infinite-dimensional Lie
algebra. The coefficients of the Lie symmetries are factors of $t$. There
are four vector fields which related to the four functions $b\left( t\right)
,~c\left( t\right) $,~$d_{1}\left( t\right) $ and $d_{2}\left( t\right) $.
They are%
\begin{equation}
\Gamma _{1}=\text{du}\frac{\text{$\mu _{1}$b}^{\prime }(t)}{(\text{$\mu _{1}$%
}+\text{$\mu _{3}$})}+\text{dx}\left( \text{b}(t)\right)   \nonumber
\end{equation}%
\begin{equation}
\Gamma _{2}=\frac{\text{du$\mu _{1}$}\left( -y\text{c}^{\prime \prime
}(t)\right) }{\text{$\mu _{5}$}^{2}(\text{$\mu _{1}$}+\text{$\mu _{3}$})}-%
\text{dx}\left( \frac{y\left( \text{c}^{\prime }(t)\right) }{\text{$\mu _{5}$%
}^{2}}\right) +\text{dy}\left( \text{c}(t)\right)
\end{equation}%
\begin{equation}
\Gamma _{3}=\frac{\text{du}\text{$\mu _{1}$}\left( -z\text{$d_{1}$}^{\prime
\prime }(t)\right) }{\text{$\mu _{6}$}(\text{$\mu _{1}$}+\text{$\mu _{3}$})}-%
\text{dx}\left( \frac{z\left( -\text{$d_{1}$}^{\prime }(t)\right) }{\text{$%
\mu _{6}$}}\right) +\text{dz}(\text{$d_{1}$}(t))
\end{equation}%
\begin{equation}
\Gamma _{4}=\frac{\text{du}\text{$\mu _{1}$}\left( yz\text{$d_{2}$}^{\prime
\prime }(t)\right) }{\text{$\mu _{6}$}(\text{$\mu _{1}$}+\text{$\mu _{3}$})}+%
\text{dx}\left( \frac{y\left( z\text{$d_{2}$}^{\prime }(t)\right) }{\text{$%
\mu _{6}$}}\right) +\text{dy}\left( -\frac{\text{$\mu _{5}$}^{2}z\text{$d_{2}
$}(t)}{\text{$\mu _{6}$}}\right) +\text{dz}(y\text{$d_{2}$}(t)),
\end{equation}%
From the vector fields we can calculate the infinitesimal transformations.

Indeed, the vector field $\Gamma _{1}$ provides the infinitesimal
transformation
\[
\bar{x}=x+\varepsilon b\left( t\right) ,~\bar{u}=u+\varepsilon \frac{\text{$%
\mu _{1}$}b^{\prime }(t)}{(\text{$\mu _{1}$}+\text{$\mu _{3}$})}.
\]%
The arbitrance of the function $b\left( t\right) $ is due to the existence
of the $u_{tx}$ term in equation (\ref{1.1}). This term means that every
function $u=u\left( t\right) $ is a solution for equation (\ref{1.1}). That
is exactly the original of the arbitrance of the above functions.

Let us now introduce the new dependent function $v=u_{t}$, such equation (%
\ref{1.1}) to be expressed as the following system%
\begin{eqnarray}
\left( \frac{\text{$\mu _{3}$}}{\text{$\mu _{1}$}}+1\right) u_{xx}u+v_{x}+%
\frac{\text{$\mu _{2}$}u_{xxxx}}{2\text{$\mu _{1}$}}+\left( \frac{\text{$\mu
_{3}$}}{\text{$\mu _{1}$}}+1\right) u_{x}{}^{2}-\frac{\text{$\mu _{4}$}%
u_{xxx}}{2\text{$\mu _{1}$}}+\frac{1}{2}\text{$\mu _{5}$}^{2}u_{yy}+\frac{1}{%
2}\text{$\mu _{6}$}u_{zz} &=&0, \\
v-u_{t} &=&0.
\end{eqnarray}

The Lie symmetries of the later system are%
\[
X_{1}=\partial _{t}~,~X_{2}=\partial _{x},~X_{3}=\partial
_{y},~X_{4}=\partial _{z}\text{ and }X_{5}=\mu _{5}^{2}z\partial _{z}-\mu
_{6}y\partial _{z}.
\]%
The vector fields $X_{1},~X_{2},~X_{3}$ and $X_{4}$ are translations and $%
X_{5}$ is a rotation in the two-dimensional place $\left\{ y,z\right\} $.
These vector fields follows from the generic vector field $\Gamma $ of
equation (\ref{1.1}) by assuming that $b\left( t\right) =b_{0},~c\left(
t\right) =c_{0}$, $d_{1}\left( t\right) =d_{1}^{0}$ and $d_{2}\left(
t\right) =d_{2}^{0}$.

Consequently, the application of these vector fields for the derivation of
similarity transformations leads to the same form of transformations as
these studied in the previous section. Nevertheless the application of the
vector fields provides less arbitrary functions, thus less calculations.

\section{Conclusion}

In this work, certain trivial reductions and hence the solutions of the
Kudryashov-Sinelshchikov equation in $(1+3)-$ dimension are listed. The
important observation is a reduction with one of the symmetry vector with
arbitrary function. The reduced equation in $(1+2)-$ dimension possesses a
series solution with required number of arbitrary constants and hence passes
the Painlev\'{e} test. 

This result can be easily connected to the parent
equation which is in $(1+3)-$ dimension and of fourth-order, hence also
fingers to the integrability of equation (\ref{1.1}). 

Further study involving computation of various nonclassical symmetry is under study.%

\section* {Acknowledgements}
 AP \& PGLL acknowledges the support of the National Research Foundation of South Africa. AP thanks the support of VRIDT through Resoluci\'{o}n VRIDT No. 096/2022 and Resoluci\'{o}n VRIDT No. 098/2022. RS acknowledges the support of Anusandhan National Research Foundation (ANRF), India for supporting the project under the Core Research Grant number: File No: CRG/2023/005418, dated 20 August, 2024. 

\section*{Conflict of Interest}
The authors declare that there are no competing interests among the authors of this work.

\end{document}